\documentclass[twocolumn,showpacs,preprintnumbers,amsmath,amssymb,superscriptaddress]{revtex4}

\usepackage{graphicx}
\usepackage{dcolumn}
\usepackage{bm}
\usepackage{enumerate}

\begin{document}


\title{Metabolic network modularity arising from simple growth processes}

\author{Kazuhiro Takemoto}
\email{takemoto@bio.kyutech.ac.jp}
\affiliation{
Department of Bioscience and Bioinformatics, Kyushu Institute of Technology, Iizuka Fukuoka 820-8502, Japan
}
\affiliation{
PRESTO, Japan Science and Technology Agency, Kawaguchi, Saitama 332-0012, Japan
}

\date{August 29, 2012}

\begin{abstract}
Metabolic networks consist of linked functional components, or modules. The mechanism underlying metabolic network modularity is of great interest not only to researchers of basic science but also to those in fields of engineering. Previous studies have suggested a theoretical model, which proposes that a change in the evolutionary goal (system-specific purpose) increases network modularity, and this hypothesis was supported by statistical data analysis. Nevertheless, further investigation has uncovered additional possibilities that might explain the origin of network modularity. In this work, we propose an evolving network model without tuning parameters to describe metabolic networks. We demonstrate, quantitatively, that metabolic network modularity can arise from simple growth processes, independent of the change in the evolutionary goal. Our model is applicable to a wide range of organisms, and appears to suggest that metabolic network modularity can be more simply determined than previously thought. Nonetheless, our proposition does not serve to contradict the previous model; it strives to provide an insight from a different angle in the ongoing efforts to understand metabolic evolution, with the hope of eventually achieving the synthetic engineering of metabolic networks.
\end{abstract}

\maketitle

\section{Introduction}
Metabolic processes are essential for physiological functions and responsible for maintaining life. As a result, it is an important and interesting topic of scientific inquiry not only for researchers in the field of basic biology but also for investigators in biotechnology and medical research. {\it Metabolism} can be defined as a series of biochemical reactions, and it is often represented as a network structure. Generically, when metabolic networks are depicted in a schematic fashion, metabolites are portrayed as nodes and reactions as edges \cite{Jeong2000,Ma2003,Arita2004}.

In the recent years, several new technologies and high-throughput methods have generated a massive quantity of genomic and metabolic network data. As a result, investigators have been actively carrying out comprehensive data analyses in an ongoing attempt to shed light on the germination and connections of metabolic networks (reviewed in \cite{Barabasi2004,Albert2005}, for example). In particular, several studies, until now, have discussed possible mechanisms underlying the evolution of metabolic networks \cite{Light2005,Pal2005,Diaz-Mejia2007} and their contribution to environmental adaptation (reviewed in \cite{Papp2008,Nam2011}).

Previous studies have often focused on dissecting the mechanism of {\it metabolic network modularity}, which, in essence, reflects the deconstruction of a network into dense, and yet, weakly interconnected subnetworks \cite{Fortunato2010,Newman2006}. Metabolic network modularity is considered one of the most vital and overarching principles governing the organization of biological networks \cite{Hartwell1999}. Previous work has demonstrated through reconstruction of the evolutionary history of metabolic networks that niche specification has reduced metabolic network modularity \cite{Kreimer2008}, and additionally revealed the effect of metabolic network modularity on the metabolic robustness \cite{Hintze2008,Holme2011}.

What has particularly captured interests in the field is the origin of network modularity. Kashtan and Alon \cite{Kashtan2005} have suggested a possible theoretic model using an evolutionary optimization algorithm based on the edge switching mechanism. In this theory, it was conjectured that modular networks spontaneously evolve when the evolutionary goal (i.e., system-specific purpose) changes over time in a manner that preserves the same sub-goals but in different permutations. In this context, an evolutionary goal can be interpreted as the variability in a species habitat. Originally, Lipson et al. suggested this evolutionary force that can lead to modularity \cite{Lispson2002}. Inspired by these studies, Parter et al. \cite{Parter2007} showed that that variability in natural habitat promotes metabolic network modularity in bacteria (i.e., the network modularity of an organism living in wider environments is higher), and they showed a mechanism possibly responsible for the change in metabolic network modularity. Moreover, Samal et al. have also derived similar conclusions on the relationship between metabolic network modularity and changes in the chemical environment, which they specifically defined as the availability and source of carbon-based molecules, using flux balance analysis \cite{Samal2011}

Nevertheless, this previously established theoretical model has limitations. It is not quantitative and is based on gene regulatory networks \cite{Kashtan2005}. Therefore, it may not be directly applicable to metabolic networks. Furthermore, the evolutionary rate of edge rewiring in metabolic networks is known to be significantly lower as compared to that in gene regulatory networks \cite{Shou2011}. Given that the previous model is constructed by the evolutionary optimization algorithm based on the edge switching mechanism, it may be difficult to completely explain the origin of metabolic network modularity through this approach. Moreover, changes in metabolic network modularity depend on growth conditions such as temperature and trophic requirements, and they are not necessarily reliant on environmental variability, i.e., changes in a natural habitat; this has been demonstrated in archaeal species \cite{Takemoto2011}. Sol\'e and Valverde \cite{Sole2008} have also questioned the view that network modularity is the result of change in evolutionary goals by citing a number of references. Using a network model, these authors claimed that such a mechanism is not required for acquiring network modularity. However, they focused on protein interaction networks, not metabolic networks, and only presented qualitative results on the origin of network modularity.

This evidence points to other possible origins of metabolic network modularity. Therefore, in this study, we focus and formulate a quantitative theoretical construct to delve into the origin of network formation. Specifically, we set out to investigate metabolic network structures from the perspective of network modularity, using a simple growing process model \cite{Takemoto2008,Takemoto2012}. We based our construct on these parameters: the network modularity measure, number of modules, and distribution of module size. By comparing the theoretical and empirical metabolic networks of 113 organisms, including 45 archaea, 60 bacteria and 8 eukaryotes (see Appendix \ref{app:organism}), we found that the simple network model could reproduce the empirical metabolic network modularity both qualitatively and quantitatively. Furthermore, juxtaposing our model against the null model confirmed the statistical significance and predictive power of our model in characterizing metabolic network modularity.

The evidence implies that metabolic network modularity can arise from simple evolutionary processes, such as the emergence of metabolic enzymes (i.e., edges) from gene duplication and horizontal gene transfer, even in absence of change in evolutionary goals, which, in previous studies, was considered an important source of metabolic network modularity \cite{Kashtan2005,Lispson2002,Parter2007,Samal2011}. In particular, metabolic network modularity seems to escalate when a shortcut path is created in an existing metabolic pathway. Such a shortcut reduces the minimum distance between two nodes in a network and can bypass a relatively short pathway. The simple relationship that exists between the network modularity and model parameters seems to strongly indicate that a universal mechanism is at work that governs changes in network modularity among different species.

The findings summarized thus far introduce the hypothesis that metabolic network modularity is possibly propelled by an alternative mechanism than what has been proposed in literature, and they delineate the simplicity and elegance that likely underlie the evolution and design of metabolic networks.

\section{A model for metabolic network formation}
\label{sec:model}
In addition to modularity, metabolic networks possess several other structural properties. For example, it is well known that metabolic networks are highly clustered and heterogeneous, or scale-free. This last feature implies that only a few hubs in the network integrate a large number of nodes, whereas most of the remaining nodes are not associated with these main hubs (reviewed in \cite{Barabasi2004,Albert2005}). These structural properties are characterized using the degree distribution and clustering coefficient \cite{Watts1998,Ravasz2003}. Degree distribution is a probability distribution of node degree, which is defined as the number of edges associated with a node, whereas the clustering coefficient indicates the edge density among the immediate neighbors of a particular node. These structural properties are dependent on the species domain such as growth temperature and habitat \cite{Zhu2005,Takemoto2007,Mazurie2010}. In this work, we propose a simple growing network model for metabolic network structure \cite{Takemoto2008,Takemoto2012}, and we proceed to show that the structural properties of the model can accurately and quantitatively predict the parameters of empirical metabolic networks. Thus, we confidently submit this model as a validated alternative model that explains the origin of metabolic network modularity.

\begin{figure}[tbhp]
\begin{center}
	\includegraphics{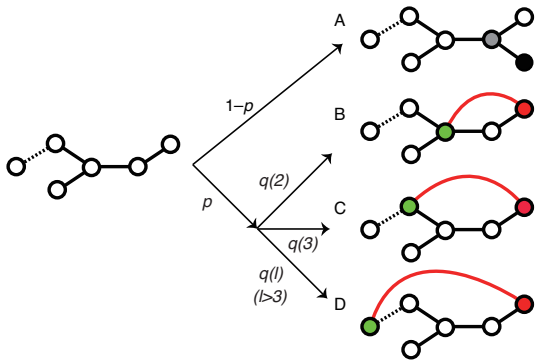}  
	\caption{(Color
 online) Schematic diagram depicting the model of growing metabolic networks.
	(A) Event I. Nodes in black and gray represent, respectively, a new node and a randomly selected existing node.
	(B), (C), and (D) Event II. Red (gray) lines correspond to new edges. Nodes in red (rightmost gray nodes) are randomly selected existing nodes, those in green (the other gray nodes) are existing nodes selected by a random walk from each red node.
	The dashed line indicates the reduced path on which nodes may exist. The new edge becomes a shortcut between the red and the green nodes. Note that the shortcut path, bypassing a path of length $l$ selected through the random walk mechanism, is accepted with the probability $q(l)$.}
	\label{fig:fig1_metabo_model_diag}
\end{center}
\end{figure}

A simple network model for unidirectional metabolic networks portray metabolites as nodes and metabolic reactions as edges; this is essentially a substrate-product relationship based on atomic tracing \cite{Arita2004}. In general, the common assumption is that metabolic networks expand by the addition or emergence of novel reactions or enzymes, resulting from evolutionary events
(see \cite{Takemoto2008,Takemoto2012} for details). Under this assumption, we can consider two situations: the case where a new reaction occurs between a new metabolite and an existing metabolite (Event I), and the case where a new reaction occurs only between existing metabolites (Event II). At each time step $t$, we assume that Event I and Event II occur with the probabilities $1-p$ and $p$, respectively (see Fig. \ref{fig:fig1_metabo_model_diag}).

In the case of Event I, a new node connects to a randomly selected node (Fig. \ref{fig:fig1_metabo_model_diag}A). That is, we can consider the model as a randomly growing tree when Event I occurs predominantly. On the other hand, in the case of Event II, a shortcut path bypasses the path of length $l$ between between one node and another (Figs. \ref{fig:fig1_metabo_model_diag}B, \ref{fig:fig1_metabo_model_diag}C, and \ref{fig:fig1_metabo_model_diag}D). This shortcut path is generated via a random walk, based on the existing network structure. The assumption implicit in random walks is that shortcut chemical reactions are likely to occur between related metabolic compounds, located in close proximity within a particular metabolic pathway (see \cite{Takemoto2008,Takemoto2012} for details). However, such random walks may be biased due to biological constraints, and we need to consider the length of the bypassed path when constructing shortcut paths.

In our previous works, we focused on the degree distribution and clustering coefficient, and demonstrated that our model is in excellent agreement with empirical metabolic networks by comparing with null models \cite{Takemoto2008,Takemoto2012}.
Especially, the degree distributions and degree exponents of model networks are almost perfectly coincident with that of empirical ones.
Moreover, we also showed that our model can predict the empirical networks in terms of degree-dependent clustering coefficients and average clustering coefficients although a null model cannot reproduce these structural parameters.
In this previous work, in addition, we only distinguished between the cases of lengths equal to 2 and the cases of lengths greater than 2 \cite{Takemoto2008} because the degree distribution is independent of the bypassed path length and the clustering coefficient is influenced only when a path of length 2 is bypassed (see \cite{Takemoto2008,Takemoto2012} for details).

As an improvement, in this work, we described the network structure in further detail, by considering a more general case where the shortcut path bypasses a path of length $l$ that is selected by the random walk mechanism and accepted with the probability $q(l)$ $(l>1)$. As an example, Figs. \ref{fig:fig1_metabo_model_diag}B and \ref{fig:fig1_metabo_model_diag}C display cases where the lengths are 2 and 3, respectively. In addition, the probability $q(l)$ is set to 0 to avoid self-loops when the length equals 0 and to prevent multiple edges when the length is 1.

In our model, we initially constructed a small metabolic network, presented as a 2-node complete graph, which evolves according to the above procedure until $N$ nodes are present. Our model assumes the parameters $p$ and $q(l)$. However, parameter tuning is not necessary; we can readily estimate these parameters by using the simple graph-theoretic statistics of an empirical metabolic network. By the definition of the model, time evolutions of the number of nodes, $N$, and the number of edges, $E$, are defined as $N=(1-p)t$ and $E=t$, respectively. Thus, the parameter $p$ is estimated as
\begin{equation}
p=1-\frac{N}{E},
\end{equation}
using $N$ and $E$ obtained from an empirical metabolic network. Assuming that metabolic networks display a sparse, or tree-like, structure (i.e., $E/N\approx 1$), the number of cycles of length $l+1$, i.e., $L_{l+1}$, increases by one with the probability $p \times q(l)$. Therefore, $L_{l+1}=p\times q(l)\times t$. As a result, the probability, $q(l)$, can be estimated as
\begin{equation}
q(l)=\frac{L_{l+1}}{E-N}
\label{eq:q_l}
\end{equation}
using $L_{l+1}$, $E$, and $N$ obtained from an empirical metabolic network. In addition, the sparse, tree-like structure of metabolic networks is essentially satisfied (see Supplemental Material).

\section{Results}
\subsection{The probability $q(l)$ decays exponentially}
As described in the previous section, the probability $q(l)$ can be estimated as $L_{l+1}/(E-N)$ from empirical metabolic networks. This poses the question of which types of functions arise from the probability $q(l)$. Given that the possibility of a metabolic reaction between a given chemical compound pair may be dependent on their structural similarities, it is expected to exponentially decay with the length $l$ of an existing metabolic pathway between chemical compounds.

\begin{figure*}[tbhp]
	\includegraphics{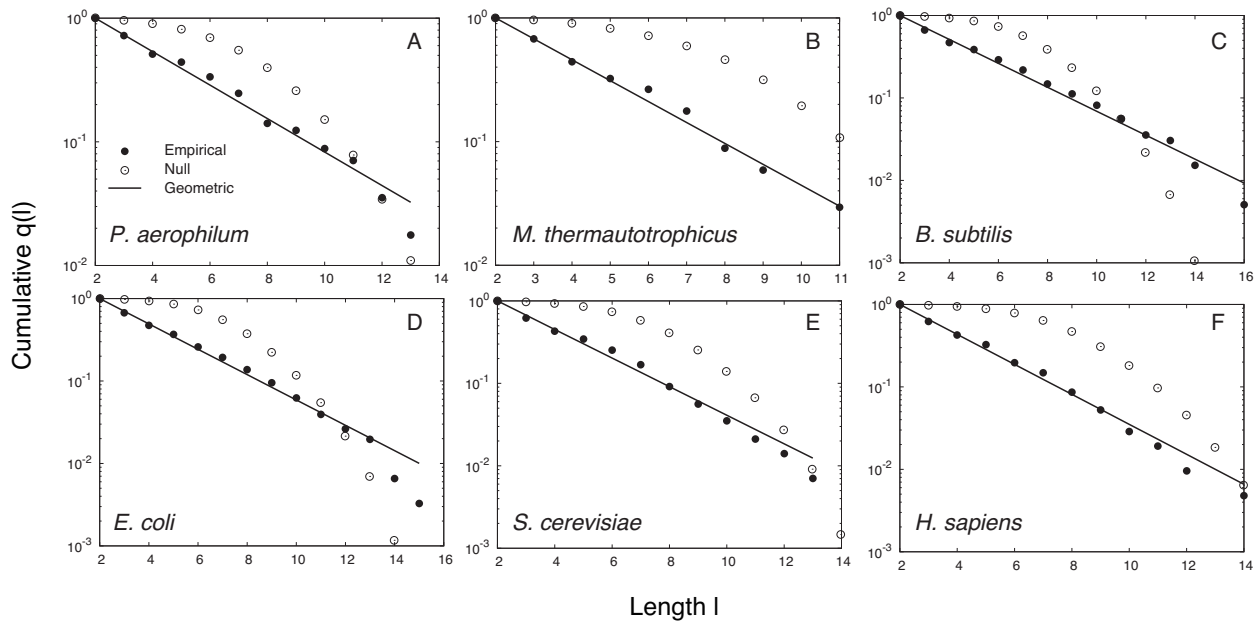}
	\caption{Probability $q(l)$ decays exponentially with the length $l$ of bypassed paths.
	The filled symbols correspond to the empirical cumulative probability distributions, $q(l)$, of {\it Pyrobaculum aerophilum} (A), {\it Methanothermobacter thermautotrophicus} (B), {\it Bacillus subtilis} (C), {\it Escherichia coli} (D), {\it Saccharomyces cerevisiae} (E), and {\it Homo sapiens} (F). Note that a cumulative distribution is defined as $P(X\geq x)$. Probability distributions from the null model (open symbols) are averaged over 300 realizations. The solid line indicates approximations of the empirical probability function $q(l)$ using the geometric distribution.}
	\label{fig:fig2_ql}
\end{figure*}

To test this assumption, we investigated the empirical metabolic networks. Figure \ref{fig:fig2_ql} shows the cumulative representation of the $q(l)$ obtained from the empirical metabolic networks of 6 representative organisms, including 2 archaea, 2 bacteria, and 2 eukaryotes. Although the above assumption of $q(l)$ is more or less intuitive, the probability $q(l)$ decreases exponentially with the length $l$, which is expected.

Because of the discrete property of length $l$ and the constraint $\sum_{l=2}q(l)=1$, the consideration of a geometric distribution may be natural to describe the exponential decay, $q(l)$:
\begin{equation}
q(l)=(1-G)^{l-2}G.
\end{equation}
Using the mean of the empirical $q(l)$, i.e., $\langle l \rangle$ (i.e., the estimated mean of bypassed paths), the parameter $G$ is estimated to be
\begin{equation}
G=\frac{1}{\langle l \rangle-1}.
\end{equation}
As shown in Fig. \ref{fig:fig2_ql}, the approximation of $q(l)$ using the geometric distribution (solid lines) is in excellent agreement with the empirical $q(l)$, as expected. However, when $\langle l \rangle=1$, the approximation using geometric distribution is not applicable because $G=\infty$. In this case, we need to consider the empirical $q(l)$.

The probability function $q(l)$ obtained from the null model (open symbols) is critically at odds with those derived from the empirical metabolic networks, showing that the exponential decay in the empirical $q(l)$ (filled symbols) is nontrivial. The null network model was generated through the randomization of empirical metabolic networks under the condition of fixed degree distributions \cite{Maslov2002} (see Appendix \ref{app:randomization} for details), and they approximately correspond to our model with simple random works (i.e., the model in which the biased random walk mechanism (probability function $q(l)$) is {\it not} explicitly considered).

\begin{figure}[tbhp]
	\includegraphics{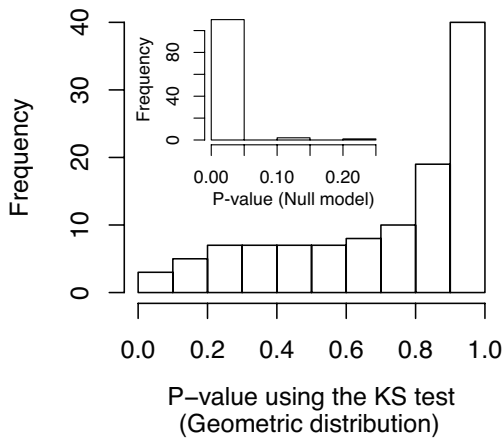}
	\caption{Geometric distributions of $P$-values of empirical and predicted models, derived using the Kolmogorov-Smirnov (KS) test.
	The KS test was performed in 113 organisms; the median $P$-value is approximately 0.8. The inset shows the $P$-value distributions of the empirical data and the null model (i.e., $q(l)$ from randomized networks) obtained using the KS test. $P$-values are averaged over 300 realizations, and the median is less than $10^{-7}$.}
	\label{fig:fig3_ql_pKS}
\end{figure}

Using the Kolmogorov-Smirnov (KS) test, the derived $P$-value (i.e., the degree of goodness-of-fit) of $q(l)$ between the empirical and estimated data also indicates that the probability distribution $q(l)$ can be approximated as the geometric distribution (Fig. \ref{fig:fig3_ql_pKS}). In contrast, the degree of goodness-of-fit of $q(l)$ between the empirical data and the null model data is close to 0 (the inset in Fig. \ref{fig:fig3_ql_pKS}). This result also indicates the significance of the exponential decay of $q(l)$ in empirical metabolic networks.

\subsection{Network modularity measure and the number of modules}
Using the estimated parameters $p$ and $q(l)$, we generated model networks as well as calculated the network modularity measure $Q$ and the number of modules $M$ using the greedy algorithm \cite{Clauset2004}. The network modularity measure $Q$ is defined as the fraction of edges that lie within modules rather than between modules, relative to that expected by chance (e.g., see Eq. (4) in \cite{Clauset2004} for the definition). Although the definition of network modularity should be further investigated, we selected this definition of network modularity based on previous studies of network modularity \cite{Kashtan2005,Parter2007,Sole2008}.

Here, we used the probability function $q(l)$, approximated using the geometric distribution. Owing to a high degree of goodness-of-fit of the geometric distribution shown in Figs. \ref{fig:fig2_ql} and \ref{fig:fig3_ql_pKS}, the network modularity measure $Q$ and the number of modules $M$ calculated from the model networks and the estimated $q(l)$ are almost similar to those calculated using the empirical $q(l)$ directly obtained from Eq. (\ref{eq:q_l}) (see Supplemental Material).

As shown in Fig. \ref{fig:fig4_compari_Q_M}, the model networks can predict the empirical $Q$ and $M$. The comparison of the network modularity measure $Q$ and the number of modules between our model and the null model (i.e., randomized networks) shows the significance of our model.

\begin{figure*}[tbhp]
	\includegraphics{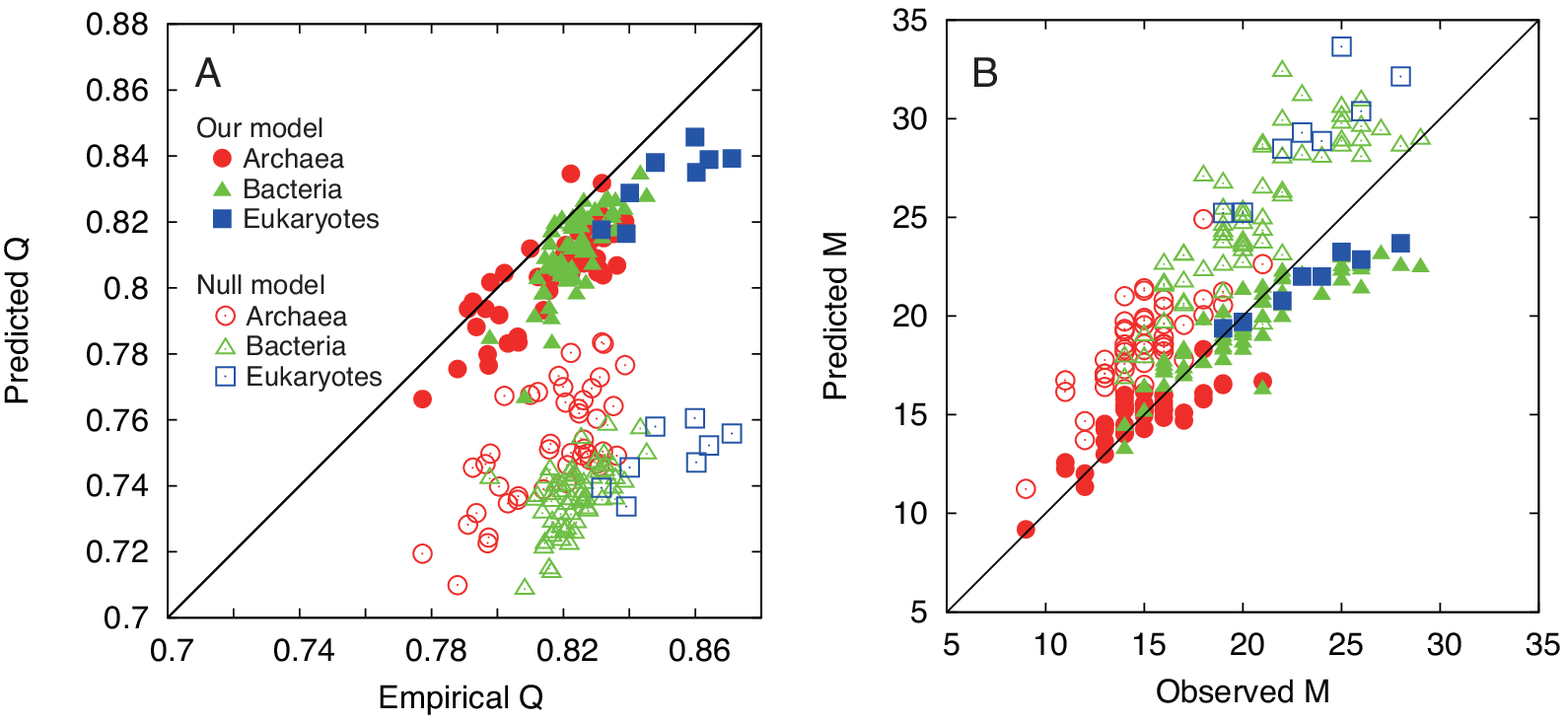}
	\caption{(Color
 online) Comparison of the network modularity measure (A) and the number of modules (B) between the empirical and model metabolic networks in 113 organisms.
	The network modularity measure and the number of modules from our model and the null model are averaged over 300 realizations.}
	\label{fig:fig4_compari_Q_M}
\end{figure*}

We evaluated the prediction accuracy of our model and the null model by using the Pearson correlation coefficient (CC) and root mean square error (RMSE) between the predicted values $x_i$ and the empirical values $y_i$: \begin{equation}
\sqrt{\frac{1}{n}\sum_{i=1}^n(x_i-y_i)^2},
\end{equation}
where $n$ is the number of samples (i.e., the number of organisms $n=113$ in this study). As can be seen in Table \ref{table:table1}, the prediction accuracy of our model is critically higher than that of the null model, indicating the significance of our model and the importance of considering the probability function $q(l)$.

\begin{table}[tbhp]
\caption{\textbf {Prediction accuracy of the network modularity measure $Q$ and the number of modules $M$, in terms of correlation coefficient (CC) and root mean square error (RMSE)}. Values with higher CC and lower RMSE indicate better predictions. Predictions that are more accurate are in bold.}
\label{table:table1}
\centering
\begin{tabular}{ccc|cc}
\hline
& \multicolumn{2}{c|}{$Q$} & \multicolumn{2}{c}{$M$} \\
\cline{2-5}
& CC & RMSE & CC & RMSE \\
\hline
Our model & {\bf 0.824} & ${\bf 1.57 \times 10^{-2}}$ & {\bf 0.939} & {\bf 1.82} \\
Null model & 0.385 & $8.08 \times 10^{-2}$ & 0.915 & 4.70 \\
\hline
\end{tabular}
\end{table}

\subsection{Module size distribution}
\begin{figure*}[tbhp]
	\includegraphics{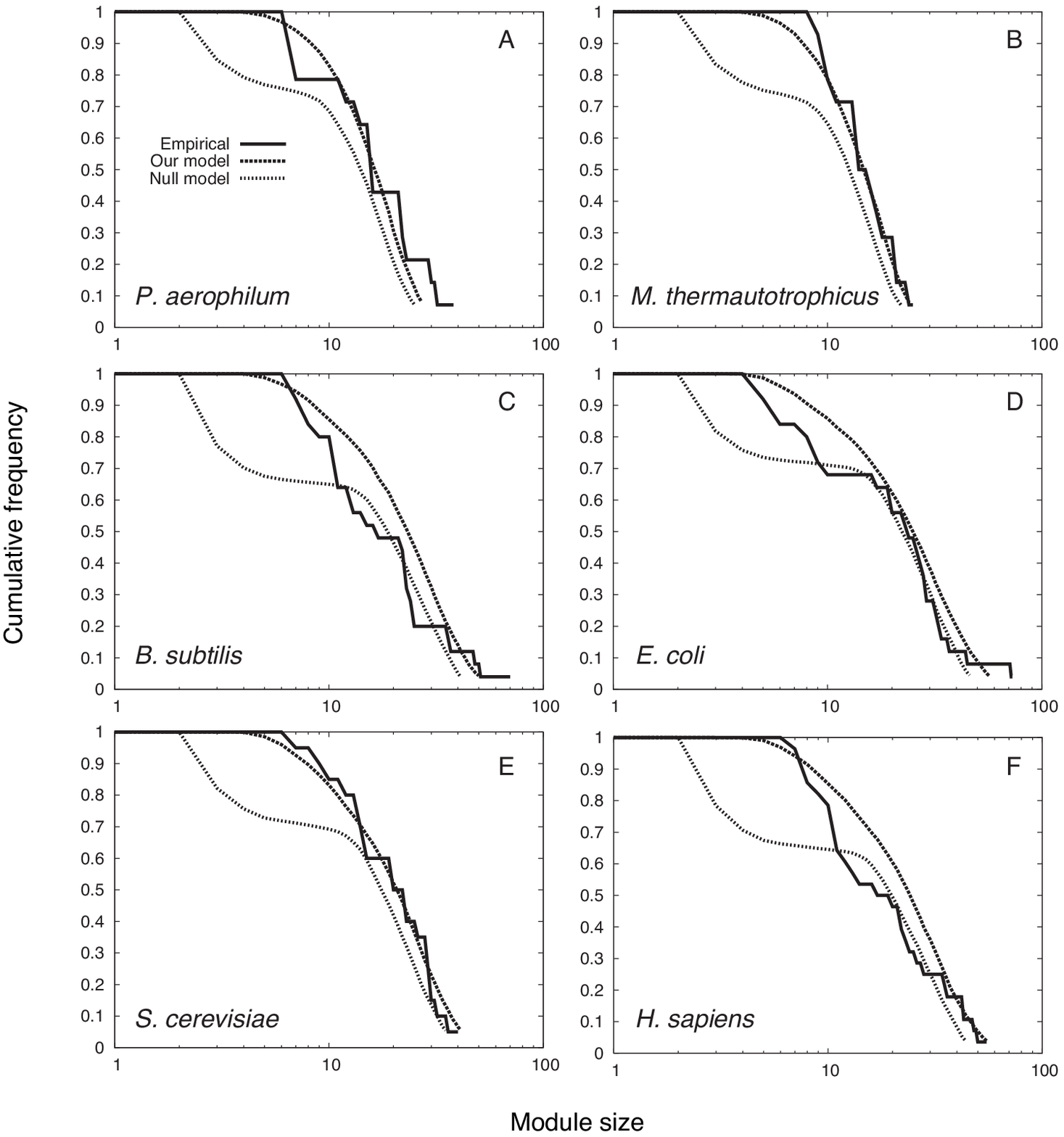}
	\caption{Cumulative distributions of module size in 6 representative organisms:
	(A) {\it Pyrobaculum aerophilum}, (B) {\it Methanothermobacter thermautotrophicus}, (C) {\it Bacillus subtilis}, (D) {\it Escherichia coli}, (E) {\it Saccharomyces cerevisiae}, and (F) {\it Homo sapiens}. Note that a cumulative distribution is defined as $P(X\geq x)$. The cumulative distributions derived from our model and the null model are averaged over 300 realizations.}
	\label{fig:fig5_module_size_dist}
\end{figure*}

To evaluate network modularity in metabolic networks in further detail, we investigated the cumulative distribution of module size, which is defined as the number of nodes in a module detected by the greedy algorithm \cite{Clauset2004} (Fig. \ref{fig:fig5_module_size_dist}). The cumulative distribution of module size repurposed using our model is in good agreement with the empirical distributions. Contrastively, the null model shows a low goodness-of-fit with the empirical data.

\begin{figure}[tbhp]
	\includegraphics{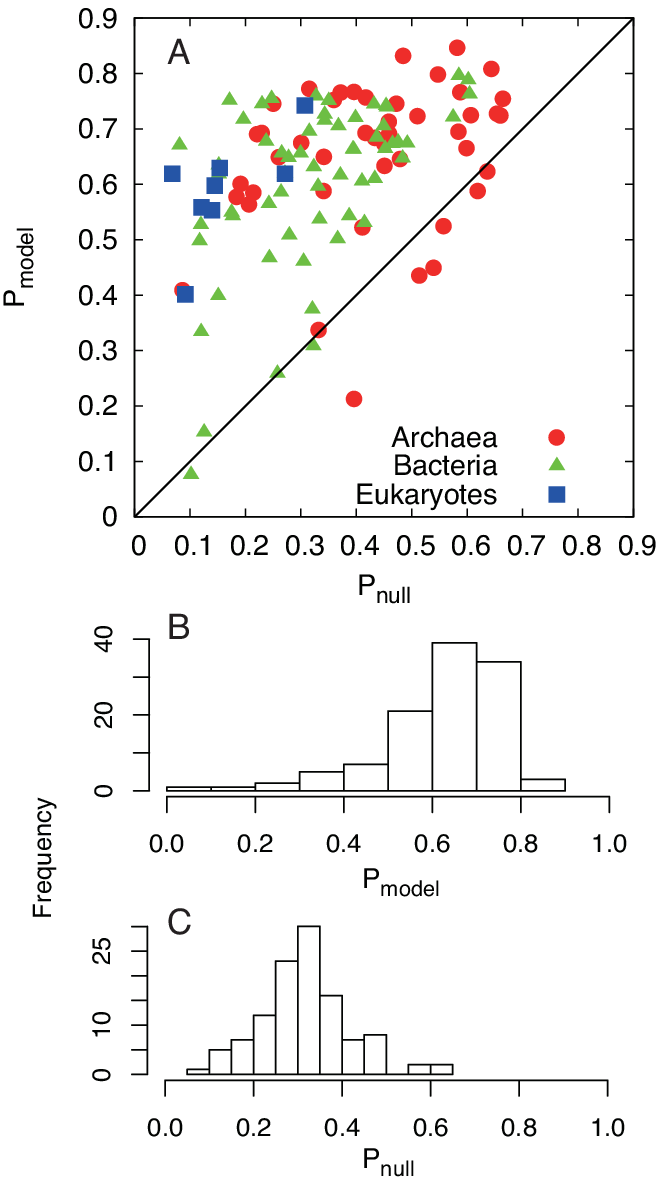}
	\caption{(Color
 online) $P$-value comparison of empirical distributions and predicted distributions using the Kolmogorov-Smirnov (KS) test.
	(A) $P_{\mathrm{model}}$ corresponds to the $P$-value when our model is applied, and $P_{\mathrm{null}}$ corresponds to that when the null model is applied. The medians of the $P_{\mathrm{model}}$ (B) and $P_{\mathrm{null}}$ (C) distributions are 0.66 and 0.31, respectively. The KS test was performed in 113 organisms. The $P$-values are averaged over 300 realizations.}
	\label{fig:fig6_all_compari_P}
\end{figure}

The comparison of the $P$-value using the Kolmogorov-Smirnov test between our model and the null model (i.e., the comparison between $P_{\mathrm{model}}$ and $P_{\mathrm{null}}$) (Fig. \ref{fig:fig6_all_compari_P}A) and the frequency distributions of $P_{\mathrm{model}}$ (Fig. \ref{fig:fig6_all_compari_P}B) and $P_{\mathrm{null}}$ (Fig. \ref{fig:fig6_all_compari_P}C) also show the significance of our model for predicting the distribution of module size.

\subsection{Simple relationship between network modularity and model parameters}
The network modularity may increase when networks are locally dense. In our model, the parameters $p$ and $G$ reflect the local denseness in networks. The parameters $p$ and $G$ are respectively the probability that short-cut paths, which reduce the minimum distance between two nodes on a network, emerge and the inverse of the average length of a pathway bypassed by a short-cut path minus 1. Therefore, the metabolic network modularity is expected to depend on the parameter $p \times G$.

\begin{figure}[tbhp]
	\includegraphics{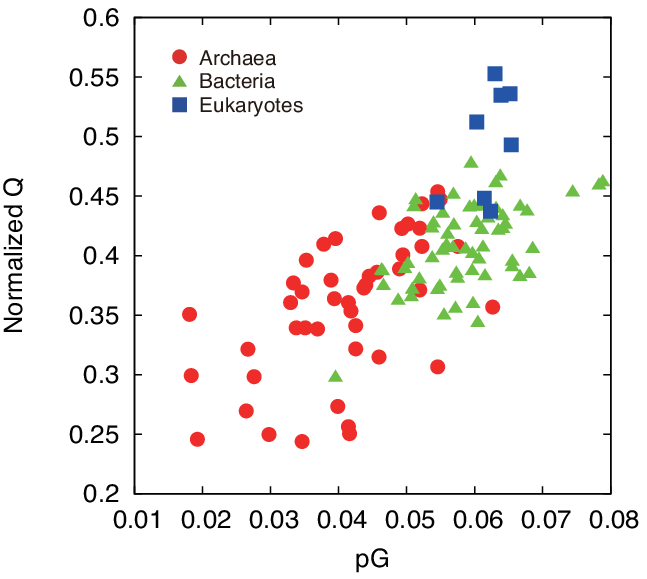}
	\caption{(Color
 online) Correlation between the normalized network modularity measure and the model parameter.
	See Appendix \ref{app:normalization} for the normalization procedure of network modularity measure.}
	\label{fig:fig7_Qm_pq}
\end{figure}

As expected, we found a correlation between the network modularity measure and model parameters $p\times G$ in the empirical metabolic networks (see Fig. \ref{fig:fig7_Qm_pq}). Note that the network modularity measure is normalized in order to avoid the effect of the numbers of nodes and edges on the network modularity measure (see Appendix \ref{app:normalization}). In particular, a linear relationship is observed across different domains, even though the organisms, including archaea, bacteria, and eukaryotes, show different network modularity measures. However, higher eukaryotes such as humans and mice seem to deviate slightly from this relationship.

This result indicates that the metabolic network modularity increases when a short-cut path occurs in existing metabolic pathways and the short-cut paths bypass a relatively short pathway, and it implies that the mechanism in the change of network modularity is universal among different species.

\section{Discussion}
To date, only qualitative models for gene regulatory networks \cite{Kashtan2005} and protein interaction networks \cite{Sole2008} have been described in literature, and no known quantitative models existed prior to our study. In this work, we propose a novel, quantitative model that proffers a unique perspective and furthers the understanding of the manner in which metabolic networks acquire modularity through simple growth processes.

As mentioned in Sec. \ref{sec:model}, we have already demonstrated that our model can successfully predict other well-known structural properties in empirical metabolic networks, such as the degree distribution, average clustering coefficient, and degree-dependent clustering coefficient \cite{Takemoto2008,Takemoto2012}. In this work, we present additional evidence to illustrate the usefulness and predictive power of our model in describing empirical metabolic networks.

Our model established that metabolic network modularity can arise through simple growth processes. This finding implies that metabolic network modularity can be acquired without changes in evolutionary goals (e.g., variability in a species habitat); this is supported by the genetic programming approach, which is based on the edge switching mechanism \cite{Kashtan2005} and metabolic network analysis \cite{Parter2007}. Consequently, this finding contradicts previously established views that metabolic network modularity arises as a result of a change in the evolutionary goal.

In particular, Figure \ref{fig:fig7_Qm_pq} suggests a common mechanism at work, governing the change in network modularity across different species, despite a few exceptions in higher eukaryotes. This result corroborates our hypothesis on the origin of metabolic network modularity. If the reverse were true, and variability in natural habitats did strongly influence metabolic network modularity, then we would not observe the simple relationship between species with vastly different habitats, as shown in Fig. \ref{fig:fig7_Qm_pq}.

These results indicate that metabolic network modularity is simply determined through selective pressure on the emergence of shortcut paths, and not upon change in the evolutionary goal. For instance, selection through temperature variations has been discussed previously \cite{Takemoto2011}. The engines of metabolic pathways consist of enzymes, which are proteins that require temperature stability to maintain structural cohesiveness. In environments with unusually high temperatures, environments lose structural integrity and are denatured and deactivated. As a result, the emergence of alternative paths may be restricted. This speculation may be also supported by a strong negative selection in genes under these extreme conditions, suggested by previous studies (e.g., \cite{Wang2002,Friedman2004}).

Previous studies firmly established that a strong negative selection is responsible for the network modularity under these extreme conditions \cite{Wang2002,Friedman2004}.

Several studies also present results that support our theory on the origin of network modularity. In \cite{Takemoto2011}, for example, it was shown that in archaea, no known correlations exist between environmental variability and metabolic network modularity. This, in part, suggests that metabolic network modularity is independent of the species habitat. According to this work, metabolic network modularity is dependent on species growth conditions such as temperature and trophic requirements and not reliant on the variability in natural habitats.

Moreover, the rate of edge rewiring in metabolic networks seems to discredit explanations of network modularity that are based on a genetic programming approach and use the edge switching mechanism \cite{Kashtan2005}. Although the rate of edge rewiring may vary slightly at each estimated divergence time, the ratio of the edge-rewiring rate in metabolic networks to that in gene regulatory networks ranges from $3.2\times 10^{-4}$ to $8.4\times 10^{-3}$ \cite{Shou2011}. Thus, it may be difficult to completely explain the origin of network modularity using the theoretical model based on edge switching.

Although other studies also have brushed upon this concept, and similarly questioned the rationale behind considering the origin of network modularity from the vantage point of a change in evolutionary goals (e.g., \cite{Sole2008}), in this work, we hone in specifically on differences in the features of biological networks that might engender network modularity. The previous model \cite{Kashtan2005} is originally based on gene regulatory networks and not on metabolic networks; therefore, assuming a relatively high rate of edge rewiring is appropriate \cite{Shou2011}. However, the difference in edge-rewiring rate between gene regulatory networks and metabolic networks does suggest a disparate origin of network modularity. Thus, our model does not necessarily contradict the previous model \cite{Kashtan2005}; rather, it is based on different parameters. We believe that the predictive power of our model proffers unique insight into the origin of network modularity and that when it is considered in context with previous models, it will serve its purpose in advancing our collective understanding of biological networks \cite{Kashtan2005,Sole2008}.

In general, our model can accurately predict instances of metabolic network modularity. However, we do observe a few exceptions on occasion. For example, Fig. \ref{fig:fig6_all_compari_P}A displays a few cases where the null model excels our model. Species such as {\it Helicobacter pylori} and {\it Archaeoglobus fulgidus}, represent unique exceptions. {\it H. pylori} is a human pathogen responsible for gastritis and peptic ulcer, whereas {\it A. fulgidus} are methanogens as well as heterotrophs and utilize diverse types of organic compounds \cite{Prokaryotes,Kristjansson1991}. In view of these exceptions, we should allow for the likelihood that our model may not accurately account for certain unique species with unusual characteristics. On the other hand, it is also possible that these data anomalies have resulted from a lack of data on metabolic networks in these particular species. It should be noted that metabolic network analyses sport limitations, especially when carrying out model validation through a comparison with empirical metabolic networks. Other limitations include the limited knowledge of metabolic reactions (i.e., missing links), reconstruction of metabolic networks based on genomic information, and failure to account for reaction stoichiometry and the direction of reaction (i.e., reversible vs. irreversible).

In our investigation, we found that $q(l)$ decreases exponentially with the length of the bypassed path, $l$. Comparison with the null model uncovered the nontrivial feature in the exponential decay, and we conjectured that this property is the direct corollary of our simple hypothesis that metabolic reactions are dependent on the structural similarities between metabolites. In fact, it is highly probable for a reaction to occur between a pair of metabolites with similar chemical structures. The empirical $q(l)$ (Fig. \ref{fig:fig2_ql}) further supports this simple hypothesis.

The indication that metabolic networks are likely formed though a simple mechanism may promise simpler methods for predicting interactions between biomolecules (i.e., link prediction, described in \cite{Lu2010}). Enzyme promiscuity \cite{Khersonsky2010}, which implies that enzymes can catalyze multiple reactions, act on more than one substrate, or exert a range of suppressions \cite{Patrick2007}, in which an enzymatic function is suppressed by overexpressing enzymes showing originally different functions, suggests the existence of many hidden metabolic reactions. These biological features in metabolism are important for designing metabolic pathways and understanding metabolic evolution; our model may be helpful for finding such hidden metabolic reactions.

The toolbox model \cite{Maslov2009} is a famous model for metabolic networks, and it assumes that the metabolic network of a given organism constitutes a subset of the universal biochemistry network, formed by the union of all the metabolites and metabolic reactions taking place in any organism. In particular, the metabolic network of an organism arises from self-avoiding random walks on the universal network. This model is in agreement with empirical metabolic networks. However, this model is restricted to prokaryotic catabolic pathways and requires universal biochemistry networks even though model networks are very slightly influenced by the topology of the universal network \cite{Pang2011}.

In contrast, our model is applicable to general metabolic networks, including both catabolic and anabolic pathways, and to prokaryotes and eukaryotes alike, although doubts remain concerning its application to higher mammalian organisms such as humans and mice. This is one of the advantages of our model. However, our model does not consider the relationships inherent in transcriptional regulations (i.e., the quadratic scaling of the number of transcription factors) explained by the toolbox model. Our model could be improved in the future by including the transcriptional regulatory machineries in metabolic networks.

The metabolic network modularity arising from simple growth processes, without depending on the change in evolutionary goals, may also be related to the neutral theory for chemical reaction networks \cite{Minnhagen2008,Bernhardsson2010,Lee2012}. This theory argues that specific features (e.g., degree distribution) in the networks are weakly correlated with a system-specific purpose, function, or causal chain. Our model does not uncover a clear correlation between the change in the evolutionary goal (purpose) and metabolic network modularity.

For simplicity, we did not consider a number of important evolutionary processes such as the node and edge deletions. For example, a theoretical study \cite{Sole2008} has shown that the deletion of edges is important for explaining network modularity in protein interaction networks. In addition to this, degree distributions might be altered because of such extinctions \cite{Enemark2007}. However, our current result, in addition to that of our previous study \cite{Takemoto2008,Takemoto2012} indicates that such mechanisms might have only negligible effects on structural properties such as network modularity, heterogeneous connectivity, and the clustering property. In metabolic networks, the low rate of edge rewiring due to evolutionary events, including the deletion of nodes, has been reported previously \cite{Shou2011}. Therefore, syntheses of chemical compounds are expected to amplify, indicating that the effect of node losses may be disregarded when considering the global tendencies of metabolic networks. However, this implies that deletions of nodes and interactions are {\it not} unnecessary. Such evolutionary mechanisms might play important roles in determining the partial (or local) interaction patterns of metabolic networks. Thus, we need to focus on such evolution processes in the future to fully comprehend the emergence of metabolic networks.

Even though limitations are inherent in our model, as with any other, we believe that our model still serves to provide unique insights into the origin, evolution, and design of metabolic networks.

\section*{Acknowledgments}
This work was supported by the JST PRESTO program. The author thanks Prof. Masanori Arita (University of Tokyo) for the helpful comments and suggestions.

\appendix

\section{Selection of organisms}
\label{app:organism}
We used previously published lists of archaeal and bacterial organisms \cite{Takemoto2007, Takemoto2011}. The metabolic networks in these datasets were well identified and available from the Kyoto Encyclopedia of Genes and Genomes (KEGG) database \cite{Kanehisa2006}. To prevent redundancies, when a bacterial species had different strains, we proceeded to use the strain whose genome was reported first as the representative strain for that species. Furthermore, we selected 8 representative eukaryotic species for whom data on metabolic networks were available in the KEGG database. Finally, 113 organisms, including 45 archaea, 60 bacteria, 8 eukaryotes, were examined (see Supplemental Material).

\section{Construction of metabolic networks}
The construction of metabolic networks follows similar protocols as those described in previous studies \cite{Takemoto2007,Takemoto2008,Takemoto2011}. We downloaded XML files (version 0.7.1) containing the metabolic network data of 113 organisms on 20 May 2011 from the KEGG database \cite{Kanehisa2006} (ftp://ftp.genome.jp/pub/kegg/xml/kgml/metabolic/organisms/. Note that beginning July 1, 2011, the KEGG ftp site is available only to paid subscribers.) Based on previous studies, the metabolic networks are represented by undirected networks (i.e., substrate graphs) in which the nodes and edges correspond to metabolites and reactions, respectively (i.e., substrate-product relationships are based on atomic mapping \cite{Arita2004}). Ubiquitous metabolites such as H$_2$O, ATP, and NADH were excluded. Moreover, the component for which the connection was the weakest (i.e., giant component) was extracted from each metabolic network to obtain more accurate calculations of network modularity.

\section{Null model}
\label{app:randomization}
Establishing the null model is important for demonstrating the statistical significance of proposed models. In this study, the null model illustrates randomized networks generated from an empirical metabolic network using the simple edge-rewiring algorithm \cite{Maslov2002}. This algorithm generates a random network by rewiring two randomly selected edges until the rewiring of all edges is completed. For example, we consider two edges: A--B and C--D, where the alphabets and lines denote nodes and edges, respectively. In this case, using the edge-rewiring algorithm we obtained the edges A--D and C--B (see \cite{Maslov2002} for details). Generally, in metabolic networks (i.e., substrate graphs) where reactions have multiple substrates and products, short cycles related to network modularity are generated as a result of network representations \cite{Takemoto2011}. Ideally, the number of short cycles should remain constant during the generation of randomized networks. However, the edge-rewiring algorithm used here does not abide by this constraint. Although the null model has such a limitation, it does not pose a significant problem in this study because the substrate graphs used are based on atomic mapping and currency metabolites are excluded. Hence, in our metabolic networks, most of the metabolic reactions (on an average, approximately 96\%) are represented as reactions with a single substrate and/or product. Therefore, short cycles generated by network representation rarely pose problems.

\section{Normalization of the network modularity measure}
\label{app:normalization}
To compare the effects of different network sizes and connectivities on metabolic network modularity, we used a previously established normalization framework \cite{Parter2007}, defined as 
\begin{equation}
\frac{Q_{\mathrm{empiri}}-Q_{\mathrm{rand}}}{Q_{\mathrm{max}}-Q_{\mathrm{rand}}},
\end{equation}
where $Q_{\mathrm{empiri}}$ is the network modularity measure of an empirical metabolic network, and $Q_{\mathrm{rand}}$ is the average network modularity value obtained from 300 randomized networks (i.e., the null model) constructed from the empirical metabolic network. Each $Q$ was calculated using the fast greedy algorithm proposed by Clauset et al. \cite{Clauset2004}. $Q_{\mathrm{max}}$ was estimated as $1-1/M$, where $M$ is the number of modules in the empirical network.


\end{document}